\documentclass[prb,aps,amssymb,twocolumn,superscriptaddress,notitlepage,longbibliography]{revtex4-1}
\usepackage{amsmath}
\usepackage{tikz}
\usepackage{graphicx}
\usepackage{subcaption}
\begin{document}

\newcommand{\half}{\frac{1}{2}}

\title{A systematic study of stacked square nets: \hspace{-.6em} from Dirac fermions to material realizations}

\author{Sebastian Klemenz}
\affiliation{Department of Chemistry, Princeton University, Princeton, NJ 08540, USA}

\author{Leslie Schoop}
\affiliation{Department of Chemistry, Princeton University, Princeton, NJ 08540, USA}
\email{lschoop@princeton.edu}

\author{Jennifer Cano}
\affiliation{Department of Physics and Astronomy, Stony Brook University, Stony Brook, New York 11974, USA}
\affiliation{Center for Computational Quantum Physics, The Flatiron Institute, New York, New York 10010, USA}
\email{jennifer.cano@stonybrook.edu}

\date{\today}

\begin{abstract}
Non-symmorphic symmetries protect Dirac line nodes in square net materials.
This phenomenon has been most prominently observed in ZrSiS.
Here, we systematically study the symmetry-protected nodal fermions that result from different ways of embedding the square net into a larger unit cell. 
Surprisingly, we find that a nonsymmorphic space group is not a necessary condition for a filling enforced semimetal: symmorphic space groups can also host nodal fermions that are enforced by band folding and electron count, that is, a combination of a particular structural motif combined with electron filling.
We apply the results of this symmetry analysis to define an algorithm, which we utilize to find square net materials with nodal fermions in specific symmorphic space groups.
We highlight one result of this search, the compound ThGeSe, which has not been discussed before in the context of nodal fermions.
Finally, we discuss how band folding can impose constraints on band connectivity beyond the connectivity of single elementary band representations.
\end{abstract}

\maketitle

\section{Motivation}

A fundamental question in solid state physics is to predict material properties from crystal structure.
Such structure-to-property relationships are useful to identify new materials with desirable physical attributes.
The motivation for our work is to predict topological semimetals, a subject of intense study in recent years. 
Topological semimetals are sought after for their extraordinary electronic and optical properties, such as gapless Fermi arcs,\cite{Wan11,Xu2015,Huang2015b,Weng2015,Xu15,Lv15,Lv15a} large magnetoresistance,\cite{Liang2015,Shekhar2015} and a giant nonlinear optical response,\cite{Morimoto2016,Wu2017} their potential use in fast optical switches or sensors,\cite{chan2017photocurrents,weber2018directly} and as a realization of the gravitational\cite{Lucas2016,Gooth2017} and chiral\cite{Xiong2015,Huang2015} anomalies.

The search for topological semimetals is facilitated by algorithms that can either predict or rule out materials based on their crystal structure, orbital content, and electron count, before computing their band structure.
The coarsest tool is a filling constraint for the space group: a filling constraint guarantees that at certain electron counts, a symmetry-preserving non-interacting ground state must be metallic. Recently, filling constraints have been computed for all space groups.\cite{Watanabe2015,Watanabe2016,Watanabe2018}
Beyond filling constraints, the connectivity of elementary band representations provides a finer tool: the elementary band representations provide a basis for all atomic band insulators, taking into account the space group, Wyckoff position, and orbital content of atoms.
Partially-filled connected elementary band representations must be metallic.
The connectivity of elementary band representations in all space groups has also recently been computed.\cite{TQC,Vergniory2017,Elcoro2017,Cano2018,Bradlyn2018}
These theoretical developments, along with the theory of symmetry indicators,\cite{Po2017,Song2018,Kruthoff2017} have lead to the discovery of many new topological semimetals.\cite{TQC,Chen2017filling,Vergniory2019,Zhang2019,Tang2019}

However, there remains a need for additional search mechanisms, both to filter through the thousands of compounds in materials databases\cite{Vergniory2019,Zhang2019,ICSD} for particularly promising candidates, as well as to predict compounds that have not been previously synthesized.
Specifically, despite the plethora of predicted materials, there are very few that display a Dirac cone with a linear dispersion persisting over a large energy range and which is isolated from other bands.

In this work, we combine a particular structural motif and orbital content with electron counting to predict nodal fermions that, in some cases, cannot be predicted from filling constraints and elementary band representations.




We focus on the two-dimensional square lattice, known in crystallography as the $4^4$ square net, with two atoms in the unit cell.\cite{Tremel1987,nuss2006geometric}
(The name is derived from each square unit cell having four corners and each site having four bonds.)
Young and Kane\cite{Young2015} proposed the square net motif as a source of nodal fermions when the two atoms in the unit cell are related by a glide symmetry in a non-symmorphic space group.
Shortly after, non-symmorphic symmetry protected Dirac cones with an extraordinarily large range of linear dispersion (2eV) were experimentally observed in the layered square net material ZrSiS.\cite{Schoop2016,neupane2016observation,xu2015two,topp2017surface}
Subsequently, non-symmorphic space groups have been extensively studied for their role in protecting nodal semimetals\cite{Bradlyn2016,Wieder2016,Wieder2016a,Young2017,Cano2019} and gapless surface states of topological insulators.\cite{Wang2016,Shiozaki2016,Wieder2018,muechler2020modular}

However, a non-symmorphic space group is not essential to protect the Dirac cones introduced by the square net.
On the contrary, different configurations of nodal fermions are possible depending on the symmetries preserved when the square net is embedded within the layered crystal structure, which is the study of the present manuscript.
Our results lead us to extend the search for Dirac materials in layered square lattices to crystals with planes containing $p4mm$ symmetry, with no need to restrict to non-symmorphic space groups.
In addition, we find that the square lattice motif can provide stronger filling constraints than can be derived from utilizing space group symmetry or elementary band representations alone.
Thus, we expect our results -- derived by ``folding'' the band structure -- are quite general and can be applied to other structure types that contain a sublattice with a smaller unit cell.

\begin{figure}[t]
	\centering
	\includegraphics[width=3.5in]{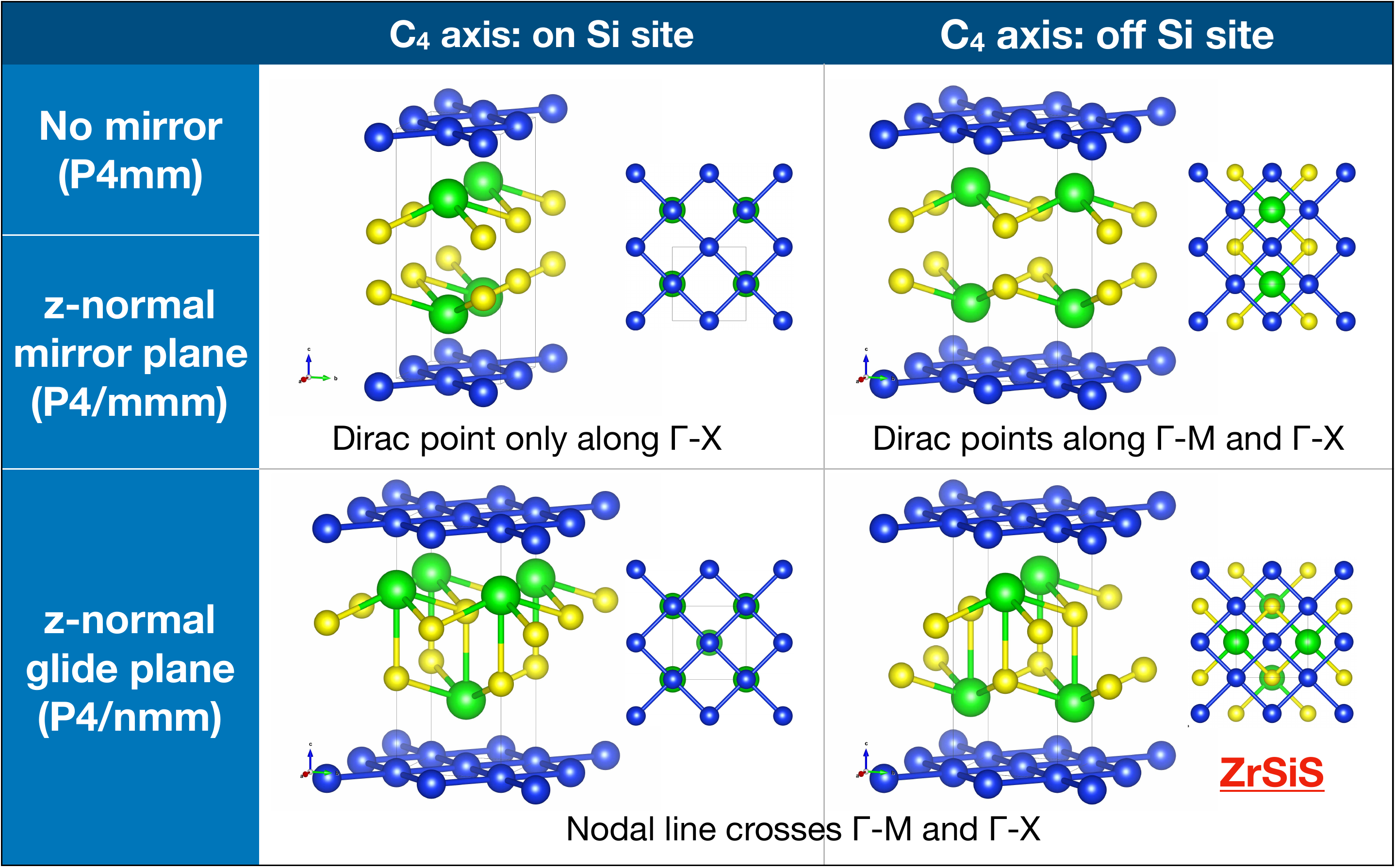}
	\caption{Nodal points and lines that result from different ways to stack a dense square net. In each box, the left picture shows an example of a crystal with the indicated symmetries, while the right pictures shows the top view.}
	\label{fig:ZrSiSvariations}
\end{figure}

We now summarize our methods and main points.
We systematically study the band structures that result from embedding a square lattice into a square unit cell that is twice as large, while preserving $p4mm$ symmetry in the plane.
Different embeddings preserve different symmetries: in particular,
only half of the $C_4$ centers of the smaller square lattice are preserved by the larger lattice, as we discuss in Sec.~\ref{sec:folding}.
Furthermore, the square lattice can be stacked in the third dimension to preserve either a $z$-normal mirror or glide symmetry, or neither; the consequences for band crossings are proven in Sec.~\ref{sec:stacking}.
The main result of this analysis is shown in Fig.~\ref{fig:ZrSiSvariations}.

Our second main result is to apply the symmetry analysis to find materials that exhibit nodal fermions in symmorphic space groups.
In Sec.~\ref{sec:space}, we list the space groups compatible with $p4mm$ layer symmetry.
Then, in Sec.~\ref{sec:materials}, we introduce an algorithm that we apply to the thousands of entries in the Inorganic Crystal Structure Database (ICSD)\cite{ICSD} in order to find material candidates.
We describe two of these candidates, ThGeSe and KCu$_2$EuTe$_4$, in detail. The former has not before been discussed in the context of nodal fermion materials.
We discuss related compounds with the same structure type as well as connections to previous work on square net materials with Dirac point and line nodes.

Our theory applies in the limit where there are no additional band inversions after folding the band structure.
We expect this limit to be valid when the spacing between layers is much larger than the atomic spacing within each layer; this is related to the tolerance factor introduced by two of us in Ref.~\onlinecite{Klemenz2019}.
Interestingly, the assumption that there are no additional band inversions after band folding leads to the prediction of Dirac points in some groups that could not be deduced from the connectivity of their elementary band representations.
We discuss this point in detail in Sec.~\ref{sec:EBR}.

Finally, we discuss the effect of spin-orbit coupling (SOC). 
Our analysis is valid in the limit of negligible SOC.
Non-negligible SOC will gap the nodal points and lines at the Fermi level.
We discuss this point in Sec.~\ref{sec:soc}.

\section{Symmetries of the square lattice}
\label{sec:symmetries}

The symmetry of a two-dimensional square lattice is described by the ``wallpaper group'' $p4mm$: it has two $C_4$ rotation centers; two parallel mirror lines in both the horizontal and vertical directions; and mirror planes along both diagonals.
Because each site is invariant under the symmetries of the point group $C_{4v}$ (also called $4mm$), the atomic orbitals transform as irreps of this group.
The group has four one-dimensional irreps (which describe, respectively, the symmetry of $p_z$, $d_{x^2-y^2}$ or $d_{xy}$ orbitals and the pseudovector $J_z$) and one two-dimensional irrep (which describes the symmetry of $p_x$ and $p_y$ orbitals; these transform identically to $d_{xz}$ and $d_{yz}$ orbitals, or as the pseudovectors $J_x$ and $J_y$);
the character table is given in Table~\ref{tab:C4v}.
The remainder of the manuscript will focus on spinless $p_x$ and $p_y$ orbitals, which describe ZrSiS and related compounds.
We assume that SOC is negligible; we return to this point in Sec.~\ref{sec:soc}.

\begin{table}
\centering
\begin{tabular}{c|c|c|c|c|c|c}
Irrep & $\mathbb{I}$ & $C_2$ & $C_4$ & $m_{100}$ & $m_{1\bar{1}0}$ & funcs. \\
\hline
$A_1$ & $1$ & $1$ & $1$ & $1$ & $1$ & $z,x^2+y^2,z^2$ \\
\hline
$A_2$ & $1$ & $1$ & $1$ & $-1$ & $-1$ & $J_z$ \\
\hline
$B_1$ & $1$ & $1$ & $-1$ & $1$ & $-1$ & $x^2-y^2$\\
\hline
$B_2$ & $1$ & $1$ & $-1$ & $-1$ & $1$ & $xy$\\
\hline
$E$ & $2$ & $-2$ & $0$ & $0$ & $0$ & $(x,y), (xz,yz) , (J_x,J_y)$ 
\end{tabular}
\caption{Character table for $C_{4v}$, reproduced from Ref.~\onlinecite{Aroyo2006}. For each irrep indicated in the first column, the characters for the group elements are listed in the middle columns. The last column indicates functions (or pseudovectors) that transform as the indicated irrep.}
\label{tab:C4v}
\end{table}

\begin{figure}
	\centering
	\includegraphics[height=1.15in]{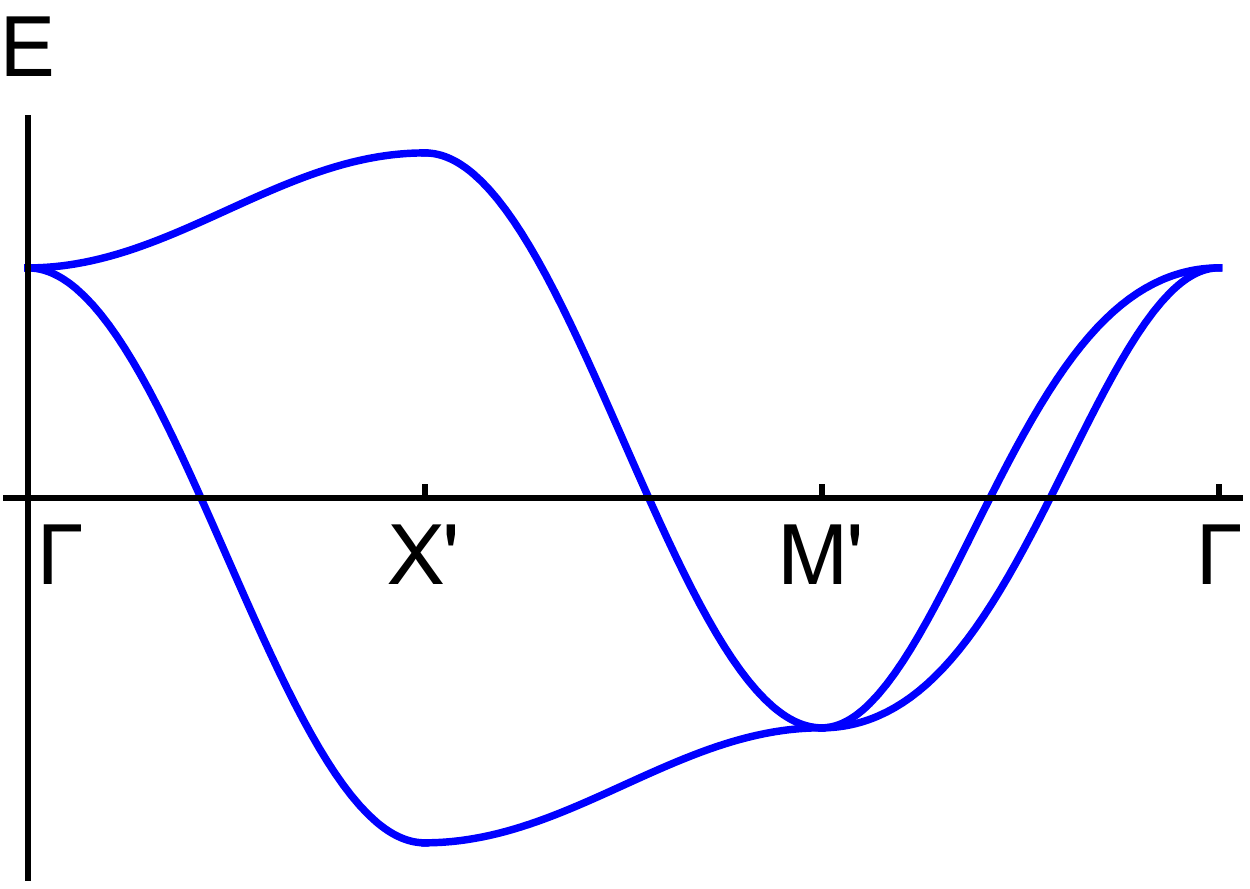}
	\includegraphics[height=1.15in]{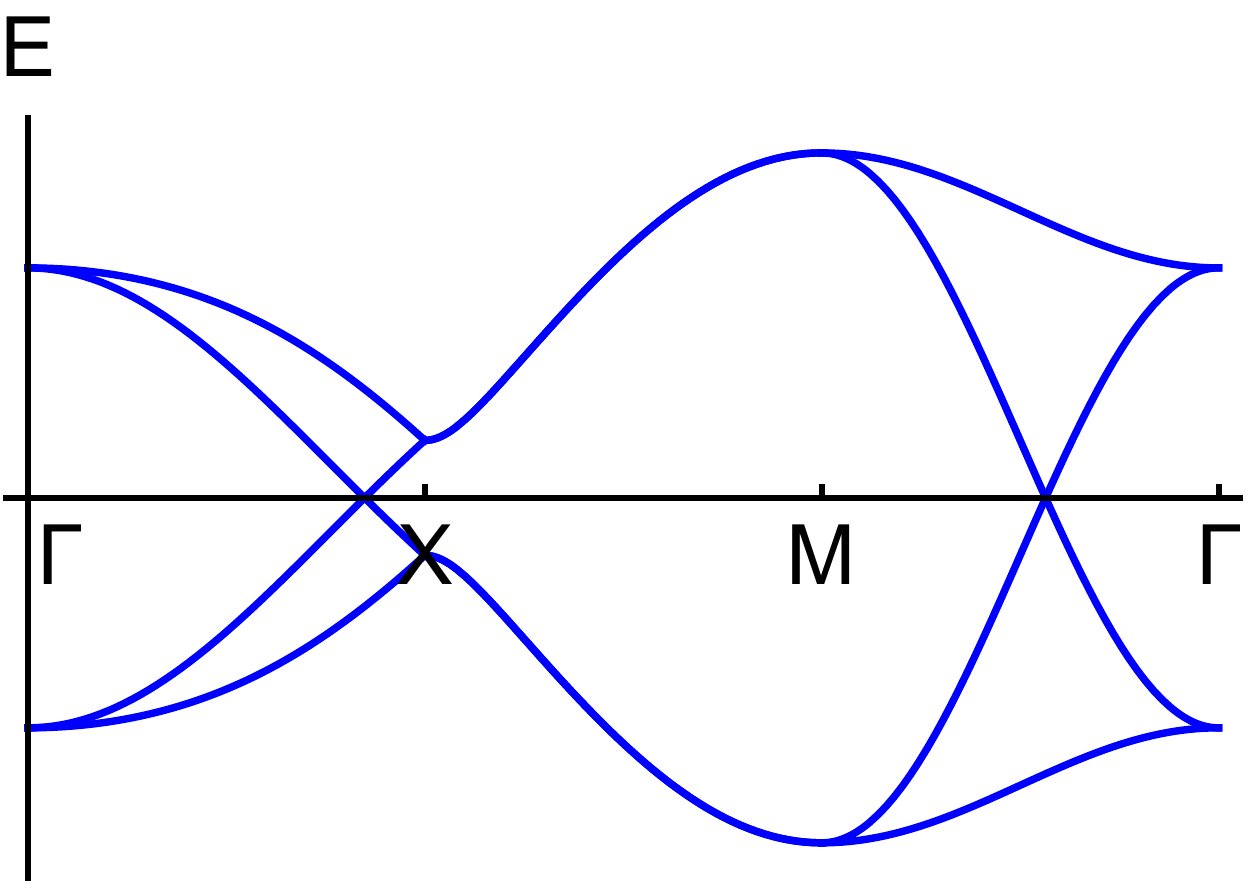}
	\caption{Band structure before (left) and after (right) band folding, with $t_\sigma=1, t_\pi = .2, t_d=.1$. The primes indicate high-symmetry points before band folding, while the unprimed points indicate high-symmetry points after band folding (see BZ in Fig.~\ref{fig:C4centers}). Which band crossings are protected depends on which symmetries are preserved in the enlarged unit cell.}
	\label{fig:SquareLatticeBands}
\end{figure}

We begin with the following minimal Hamiltonian (only nearest- and next-nearest-neighbor hopping), written in the basis of $p_x$ and $p_y$ orbitals:
\begin{align}
H_0 = & \begin{pmatrix} t_\sigma\cos k_x - t_\pi \cos k_y & -2t_d\sin k_x \sin k_y \\  -2t_d\sin k_x \sin k_y & t_\sigma \cos k_y - t_\pi \cos k_x \end{pmatrix} 
\label{eq:ham0}
\end{align}
where $t_\sigma(t_\pi)$ describes $\sigma$-bonds ($\pi$-bonds) between nearest neighbors and $t_d$ parameterizes the hopping strength diagonally across the square plaquettes.
The spectrum is shown in Fig.~\ref{fig:SquareLatticeBands}; the primes on the labels of the high-symmetry points serve to distinguish them from the folded Brillouin zone (BZ), which we will consider shortly.
Since the $\Gamma$ and $M'=(\pi,\pi)$ points are invariant under the full point group symmetry ($C_{4v}$), the bands are two-fold degenerate at those points, while there is no degeneracy at $X=(\pi,0)$ because it is only invariant under $C_{2v}$, which has no two-dimensional irreps.
This symmetry analysis can be looked up using the BANDREP application on the Bilbao Crystallographic Server.\cite{TQC,Vergniory2017,Elcoro2017}

Eq.~(\ref{eq:ham0}) is the shortest-range Hamiltonian which has only symmetry-required degeneracies (if $t_d=0$ then the bands along $\Gamma-M'$ are degenerate).
The addition of longer ranger hopping terms will deform the spectrum but cannot break the degeneracies at $M'$ and $\Gamma$.

\section{Band folding}
\label{sec:folding}

\begin{figure}
	\centering
	\includegraphics[height=1.2in]{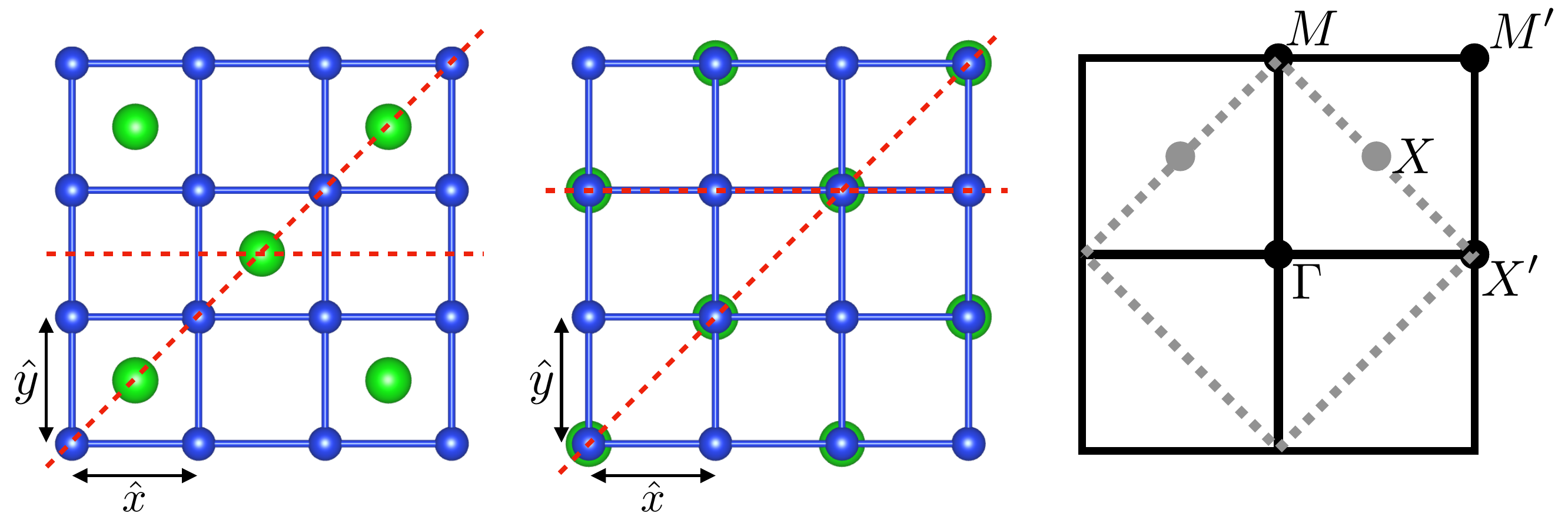}
	\caption{There are two ways to stack the larger square net (green) below a denser square net (blue) while preserving $C_4$ symmetry. In both the left and center figures, there is a $C_4$ center located on the green sites. In the left figure, the $C_4$ center on the blue sites is broken, while in the center figure the $C_4$ center in the center of the blue squares is broken.
The red dashed lines show the mirror lines.
Both arrangements yield the same folded BZ, shown on the right: the dashed line indicates the folded BZ, while each quadrant of the original BZ is outlined in solid black.}
	\label{fig:C4centers}
\end{figure}

We now consider a crystal that contains a second layer, consisting of a $\sqrt{2} \times \sqrt{2}$ lattice.
The original layer is now referred to as a dense square net.
There are two possible stacking arrangements of the two lattices that preserve a $C_4$ symmetry, which are depicted in Fig.~\ref{fig:C4centers}.
If the atoms on the larger sublattice contribute negligibly to the bands at the Fermi level, then the leading order effect of enlarging the unit cell is to fold the band structure of the original atoms, as shown on the right side of Fig.~\ref{fig:SquareLatticeBands}.
We now ask whether the two band crossings in the folded band structure (along $\Gamma-X$ and $\Gamma-M$) are symmetry protected.
The answer depends on which symmetries are preserved when the unit cell is enlarged: 
as shown in Fig.~\ref{fig:C4centers}, each of the two possible stacking arrangements preserves exactly one $C_4$ center in the unit cell, which is either located on one of the atoms (``onsite'') or on the plaquette center (``offsite'') in the original unit cell.

In the next two subsections, we prove that when the onsite $C_4$ symmetry is preserved, only the crossing along $\Gamma-X$ survives, while if the offsite $C_4$ center is preserved, both the crossings along $\Gamma-M$ and $\Gamma-X$ are protected. This result is summarized in the top row of Fig.~\ref{fig:ZrSiSvariations}.

\begin{figure}
	\centering
	\includegraphics[height=1in]{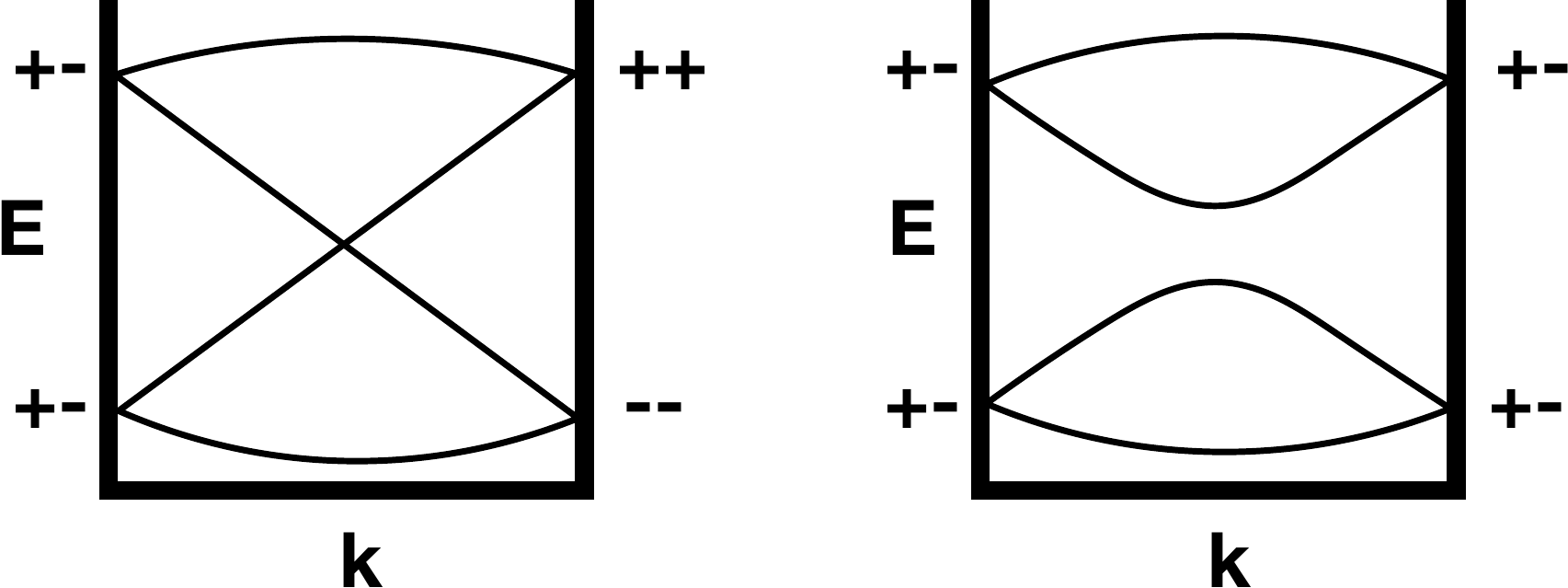}
	\caption{Two possibilities for paired mirror eigenvalues along a mirror-invariant line. Left: if at one endpoint both eigenstates within a pair have opposite mirror eigenvalues, while at the other endpoint both eigenstates within a pair have the same mirror eigenvalue, then a band crossing is required along the line. Right: if at both endpoints both eigenstates within a pair have opposite mirror eigenvalues, then the crossing is generically avoided.}
	\label{fig:bandcrossing}
\end{figure}

To prove this, we utilize the following fact (shown in Fig.~\ref{fig:bandcrossing}) that applies to a four-band model: if at both endpoints of a mirror-invariant line, eigenstates within a degenerate pair of bands always have opposite mirror eigenvalues, then there is generically an avoided crossing along the line.
On the other hand, if at one endpoint, bands within a degenerate pair have the same mirror eigenvalue, while at the other endpoint they have opposite mirror eigenvalues, a band crossing is required.
These facts are readily established by testing all possible mirror eigenvalue arrangements and noting that in the former case the bands that cross have the same mirror eigenvalue, while in the latter case there is always a crossing between bands with opposite mirror eigenvalues.

\subsection{Band crossing along $\Gamma-X$}
\label{sec:GXcrossing}

To be concrete, we choose a coordinate system so that the original lattice sites are located at $n_1 \hat{x} + n_2\hat{y}$, where $n_{1,2} \in \mathbb{Z}$, as shown in Fig.~\ref{fig:C4centers}.
Then the segment $\Gamma-X$ in the folded BZ is given by $(k,k)$, with $0\leq k\leq \frac{\pi}{2}$.
It is invariant under the mirror symmetry $m_1 : (k_x,k_y) \mapsto (k_y,k_x)$.
Both arrangements in Fig.~\ref{fig:C4centers} are invariant under $m_1$, which is shown in real space by the diagonal red dashed line.

Two distinct points in the original BZ map to each point in the folded BZ.
In particular, $\Gamma'$ and $M'$ both map to $\Gamma$, while $\pm \frac{1}{2}M'$ both map to $X$.
We now consider their $m_1$ eigenvalues.
At $\Gamma'$, two bands are degenerate before band folding. Since the $m_1$ symmetry exchanges the $p_x$ and $p_y$ orbitals, the degenerate bands at $\Gamma'$ must have opposite $m_1$ eigenvalues.
The same holds for $M'$.

In contrast, each band at $\frac{1}{2}M'$ has a degenerate partner at $-\frac{1}{2}M'$, related by time-reversal symmetry.
Since the $m_1$ eigenvalues are real and since $m_1$ commutes with time-reversal, the degenerate bands at $\pm \frac{1}{2}M'$ have the same $m_1$ eigenvalue.

Thus, we conclude that in the folded BZ, each eigenstate at $\Gamma$ has a degenerate partner with the opposite $m_1$ eigenvalue, while each eigenstate at $X$ has a degenerate partner with the same $m_1$ eigenvalue.
This is exactly the situation depicted on the left-hand side of Fig.~\ref{fig:bandcrossing}: hence, there is a required band crossing along this line.
Since both stacking configurations in Fig.~\ref{fig:C4centers} have the same $m_1$ symmetry, this band crossing is symmetry protected in both cases.


\subsection{Band crossing along $\Gamma-M$}

Maintaining the same coordinate system as in the previous section, the line $\Gamma-M$ in the folded BZ is given by $(0,k_y)$, where $0\leq k_y \leq \pi$.
However, the situation along $\Gamma-M$ is different than along $\Gamma-X$ because the two different stacking configurations in Fig.~\ref{fig:C4centers} obey different mirror symmetries that leave the $\Gamma-M$ line invariant, which is indicated by the different positions of the horizontal red dashed mirror line.
The center configuration in Fig.~(\ref{fig:C4centers}) is invariant under the mirror 
\begin{equation}
m_x: (x,y) \mapsto (-x,y),
\end{equation}
while the left configuration has a different mirror plane, 
\begin{equation}
\tilde{m}_x : (x,y) \mapsto (-x+1,y).
\end{equation}
(The original lattice is invariant under both $m_x$ and $\tilde{m}_x$, which are related to each other by a translation by $\hat{x}$, one of the original lattice vectors. When the unit cell is enlarged, translation by $\hat{x}$ is no longer a lattice vector, and it follows that only one of $m_x$ and $\tilde{m}_x$ remains a symmetry of the enlarged cell.)
Both $m_x$ and $\tilde{m}_x$ have the same action in momentum space, mapping $(k_x,k_y) \mapsto (-k_x,k_y)$.
Yet, we will show that their different actions in real space determines whether or not the band crossing along $\Gamma-M$ is protected.

Recall that two distinct points in the first BZ of the original lattice map to each point in the folded BZ.
In particular, $\Gamma'$ and $M'$ both map to $\Gamma$, while $X'$ and $C_4X'$ both map to $M$.

Since $m_x$ and $\tilde{m}_x$ are symmetries of the original lattice, we find their eigenvalues at particular points in the BZ before band folding and deduce that even after band folding, the eigenvalues will be unchanged.
We can then write $\tilde{m}_x = t_x m_x = m_x t_x^{-1}$, where $t_x$ is a translation by $\hat{x}$.
Acting on a Bloch wave function, $u_\mathbf{k}$,
\begin{equation}
\tilde{m}_x u_\mathbf{k} = m_x t_x^{-1} u_\mathbf{k} = e^{ik_x} m_x u_\mathbf{k}
\label{eq:mxeigs}
\end{equation}
Thus, we can determine the $\tilde{m}_x$ eigenvalues from those of $m_x$.

As before, the bands at $\Gamma$ and at $M'$ are degenerate before band folding.
Since $p_x$ and $p_y$ orbitals have opposite eigenvalues under $m_x$, the degenerate eigenstates at $\Gamma$ have opposite $m_x$ eigenvalues, as do the degenerate eigenstates at $M'$.
From Eq.~(\ref{eq:mxeigs}), we deduce that the same is true for the $\tilde{m}_x$ eigenvalues.

In contrast, each eigenstate at $X'$ has a degenerate partner at $C_4X'$.
We would like to know if these degenerate eigenstates have the same or opposite $m_x$ eigenvalue.
To do this, we utilize the commutation relation
$ m_x C_4 = C_4^{-1} m_x $.
Then suppose that $m_x u_{X'} = \lambda u_{X'}$, where $u_{X'}$ is a Bloch eigenstate at $X'$.
Then 
\begin{equation}
m_x (C_4 u_{X'}) = C_4^{-1} m_x u_{X'} = \lambda C_4^{-1} u_{X'} = - \lambda (C_4 u_{X'}),
\label{eq:mxeigsXp}
\end{equation}
where the last equality follows because $C_2 = -1 $ when acting on $p_{x,y}$ orbitals.
From Eq.~(\ref{eq:mxeigsXp}), we deduce that each Bloch eigenstate at $X'$ has the opposite $m_x$ eigenvalue as its degenerate partner at $C_4X'$.
After band folding, this puts us in the situation depicted on the right side of Fig.~\ref{fig:bandcrossing}: at both ends of the $\Gamma-M$ line segment, the degenerate pairs of bands have opposite $m_x$ eigenvalues and hence the band crossing is not symmetry protected.

However, from Eq.~(\ref{eq:mxeigs}), $\tilde{m}_x u_{X'} = -\lambda u_{X'}$ and $\tilde{m}_x (C_4 u_{X'}) = -\lambda (C_4 u_{X'})$, that is, each eigenstates at $X'$ has the same $\tilde{m}_x$ eigenvalue as its degenerate partner at $C_4 X'$.
After band folding, then, the situation is described by the left side of Fig.~\ref{fig:bandcrossing}: at one end of the $\Gamma-M$ line segment, the degenerate pairs of bands have opposite $\tilde{m}_x$ eigenvalues, while at the other end, they have the same $\tilde{m}_x$ eigenvalues.
It follows that a symmetry-protected band crossing is required.

To summarize: we have proven that when the onsite $C_4$ symmetry is preserved after enlarging the unit cell, the crossing along $\Gamma-M$ will generically gap, while when the offsite $C_4$ symmetry is preserved, the crossing along $\Gamma-M$ is symmetry protected.

\section{Stacked layers in three dimensions}
\label{sec:stacking}

We now consider stacking 2D layers by translating them in the $\hat{z}$ direction.
If layers with $p4mm$ symmetry are stacked by translating in the $\hat{z}$ direction, but with no additional symmetry, then the crystal is in the space group $P4mm$ (SG 99). 
(The capital $P$ indicates a space group in 3D, while the lowercase $p$ indicates a set of 2D symmetries.)
Since each $\hat{z}$-normal 2D slice in $P4mm$ has the symmetry of $p4mm$, which we analyzed in Sec.~\ref{sec:folding}, and there are no additional symmetries in $P4mm$ to impose extra constraints, the band crossings follow from Sec.~\ref{sec:folding} in this case.

If, in addition to the $\hat{z}$ translation symmetry, there is a mirror symmetry,
\begin{equation}
m_z: (x,y,z) \mapsto (x,y,-z),
\end{equation}
then the crystal is in space group $P4/mmm$ (SG 123).
However, since $p_x$ and $p_y$ orbitals are invariant under $m_z$, the $m_z$ symmetry acts like an identity operator and does not protect any additional band crossings. 
This explains why Fig.~\ref{fig:ZrSiSvariations} does not distinguish between $P4mm$ and $P4/mmm$.
The band crossings are identical to the analysis in Sec.~\ref{sec:folding}.

The third possibility (which describes the symmetry of ZrSiS\cite{Schoop2016}) is more interesting: if in addition to the $\hat{z}$ translation symmetry, there is no $m_z$, but there is a glide symmetry, then additional band crossings can be protected.
In the basis introduced in Sec.~\ref{sec:GXcrossing} and shown in Fig.~\ref{fig:C4centers}, the glide symmetry acts by:
\begin{equation}
g_z: (x,y,z) \mapsto (x+1, y, -z)
\end{equation}
However, after band folding, a translation by $\hat{x}$ is no longer a symmetry of the lattice, but is instead a fraction of a lattice translation; in the primitive basis of the larger $\sqrt{2} \times \sqrt{2}$ lattice,
\begin{equation}
g_z: (x_p, y_p, z) \mapsto (x_p +\frac{1}{2},y_p + \frac{1}{2}, -z)
\label{eq:gzprim}
\end{equation}
Consequently, $g_z$ is truly a glide symmetry of the enlarged lattice (it is a mirror symmetry when regarded with respect to only the smaller lattice).
As can be seen in the bottom row of Fig.~\ref{fig:ZrSiSvariations}, this symmetry results when the $\sqrt{2}\times\sqrt{2}$ layers are rotated by $90^\circ$ relative to each other.

Young and Kane elegantly explained this case in Ref.~\onlinecite{Young2015}.
For our purposes, there are two main results: 1) both the band crossings along $\Gamma-X$ and $\Gamma-M$ are protected by the glide symmetry and 2) the protected band crossings are part of a line node that lies in the $k_z=0$ plane. For completeness, we briefly rederive these results.

\subsection{Glide symmetry protects band crossings along $\Gamma-X$ and $\Gamma-M$}
In the primitive basis, $g_z^2$ is equal to a translation by $\hat{x}+\hat{y}$.
Since lattice translations act on the wavefunction by imposing a phase, the eigenvalues of $g_z$ are $\pm e^{-i(k_x + k_y)/2}$.
Thus, a band with $\pm 1$ eigenvalue at $\Gamma$ has $\mp i$ eigenvalue at $X$ and $\mp 1$ eigenvalue at $M$.
Crystal symmetry requires that the two-fold degenerate bands at $\Gamma$ have the same $g_z$ eigenvalue, while the two-fold degenerates bands at $X$ and $M$ are pairs with opposite $g_z$ eigenvalues; the symmetry eigenvalues can be found using the BANDREP application.\cite{TQC,Vergniory2017,Elcoro2017} (At $X$, since the bands have imaginary eigenvalues, time reversal symmetry also requires that bands with $\pm i$ eigenvalues are degenerate.)
Thus, from the analysis leading to Fig.~\ref{fig:bandcrossing}, we deduce that band crossings between $\Gamma-X$ and $\Gamma-M$ are required.

\subsection{Crossings protected by glide symmetry are part of in-plane nodal lines}

To prove that each band crossing is part of a degenerate line node, we consider a local Hamiltonian near the band crossing, restricted to the $k_z=0$ plane.
Since the bands that cross have opposite $g_z$ eigenvalues, it must be that $g_z$ is proportional to $\sigma_z$, in the basis of the two bands.
Further, since $g_z$ leaves each point in the $k_z=0$ plane invariant, the $g_z$ operator must commute with the Hamiltonian in this plane.
It follows that the Hamiltonian, up to an overall constant, must also be proportional to $\sigma_z$, that is,
$H=h_0(k_x,k_y) \sigma_0 + h_z(k_x,k_y)\sigma_z$, where $h_z(k_x,k_y)=0$ at the band crossing.
Since $h_z(k_x,k_y)$ is a function of two variables, the equation $h_z(k_x,k_y)=0$ has solutions that are lines, not points; thus, the band crossings must be part of a line of degeneracies in the $k_z=0$ plane satisfying $h_z(k_x,k_y)=0$.

Geometrically, we can further deduce that the line node must circle the $\Gamma$ point; there is no other way to draw a line in the plane that yields crossings along $\Gamma-X$ and $\Gamma-M$ but not along $M-X$.
This fact is illustrated by the ab initio calculations for ZrSiS.\cite{Schoop2016}


\section{Space groups}
\label{sec:space}

In the previous section, we showed that the space groups $P4mm$, $P4/mmm$ and $P4/nmm$ exhibit symmetry protected Dirac points or line nodes when the crystal structure consists of layered square nets of different sizes; the results are summarized in Fig.~\ref{fig:ZrSiSvariations}.

We now seek other space groups that are compatible with the stacked square lattice motif and which have enough symmetry to protect Dirac point and line nodes.
By compiling a list of space groups, we can systematically search for materials that will realize these features.
In particular, we can apply the tolerance factor developed in Ref.~\onlinecite{Klemenz2019} to find promising Dirac semimetal materials.

Our procedure is to find the ``layer groups'' -- symmetries of two-dimensional systems embedded in three dimensions -- that yield Dirac point and line nodes following the logic in Sec.~\ref{sec:folding} and then search for space groups that contain the desired layer group as a subgroup. The results are tabulated in Table~\ref{tab:spacegroups}.

Our first observation is that the four layer groups $p4mm$, $p422$, $p\bar{4}2m$, and $p\bar{4}m2$ act identically on spatial points in 2D: Table~\ref{tab:layergroups} shows the generators of these groups explicitly.
The groups differ by their action in the third dimension.
However, since $p_x$ and $p_y$ orbitals transform like vectors in 2D, the four layer groups are indistinguishable when constrained to $p_x$ and $p_y$ orbitals.
Thus, the arguments in the previous section regarding protected band crossings in $P4mm$ also apply to $p4mm$, $p422$, $p\bar{4}2m$, and $p\bar{4}m2$.

\begin{table}
\centering
\begin{tabular}{c|c|c|c|c}
2D Symmetry action & $p4mm$ & $p422$ & $p\bar{4}2m$ & $p\bar{4}m2$  \\
\hline
$(-y,x)$ & $C_{4z}$ & $C_{4z}$ & $\bar{C}_{4z}^{-1}$ & $\bar{C}_{4z}^{-1}$ \\
$(-x,y)$ & $m_x$  & $C_{2y}$  & $C_{2y}$ & $m_x$ \\
$(y,x)$ & $m_{1\bar{1}0}$  & $C_{2,110}$ & $m_{1\bar{1}0}$ & $C_{2,110}$ \\
\end{tabular}
\caption{Generators of the layer groups $p4mm$, $p422$, $p\bar{4}2m$ and $p\bar{4}m2$.
When restricted to two dimensions, the four layer groups are identical: each group has a generator (listed in the column corresponding to the group) that maps the point $(x,y)$ to one of the points in the first column.
Since $p_x$ and $p_y$ orbitals transform like vectors in 2D, the groups act identically on these orbitals.
The notation and group action were obtained from the LAYER application on the Bilbao Crystallographic Server (BCS).\cite{Aroyo2006,Aroyo2006b,Aroyo2011}}
\label{tab:layergroups}
\end{table}

In addition, the arguments in Secs.~\ref{sec:folding} and~\ref{sec:stacking} apply to $p4/mmm$, which has one more generator (the inversion symmetry operation) compared to $p4mm$, since the extra generator does not change the band degeneracies.

Finally, the layer groups with a screw or glide symmetry will always protect a Dirac point or line node following Ref.~\onlinecite{Young2015}. The layer groups with a screw or glide symmetry and a $C_4$ rotation or rotoinversion axis are: $p4/m$, $p4/n$, $p42_12$, $p4bm$, $p\bar{4}2_1m$, $p\bar{4}b2$, $p4/nbm$, $p4/mbm$, and $p4/nmm$.

\begin{table}
\centering
\begin{tabular}{cc}
Layer group & Space groups\\
\hline
$p4mm$ & $P4mm (99)$, $I4mm (107)$, $P4/mmm (123)$,\\
	& $P4/nmm (129)$, $I4/mmm (139)$ \\
\hline
$p422$ & $P4/mcc (124)$, $P4/nnc (126)$\\
\hline
$p\bar{4}2m$ & $P\bar{4}2m (111)$, $I\bar{4}2m (121)$, $P4_2/mcm (132)$,\\
	& $P4_2/nnm (134)$ \\
\hline
$p\bar{4}m2$ & $P\bar{4}m2 (115)$, $I\bar{4}m2 (119)$, $P4_2/mmc (131)$, \\
	& $P4_2/nmc (137)$, $I4_1/amd (141)$ \\
\hline
$p4/mmm$ & $P4/mmm (123)$, $I4/mmm (139)$\\
\hline
\hline
$p4/n$ & $P4/nnc(126)$, $P4/ncc(130)$ \\
\hline
$p42_12$ & $P4/mnc(128)$, $P4/ncc(130)$\\
\hline
$p4bm$  & $P4bm(100)$, $I4cm(108)$, $P4/nbm(125)$ \\
	& $P4/mbm(127)$, $I4/mcm(140)$ \\
\hline
$p\bar{4}2_1m$ & $P\bar{4}2_1m(113)$, $I\bar{4}2m(121)$, $P4_2/mnm(136)$\\
	& $P4_2/ncm(138)$ \\
\hline
$p\bar{4}b2$ & $P\bar{4}b2(117)$, $I\bar{4}c2(120)$, $P4_2/nbc(133)$ \\
	& $P4_2/mbc(135)$, $I4_1/acd(142)$\\
\hline
$p4/nbm$ & $P4/nbm(125)$, $I4/mcm(140)$\\
\hline
$p4/mbm$ & $P4/mbm(127)$, $I4/mcm(140)$ \\
\hline
$p4/nmm$ & $P4/nmm(129)$, $I4/mmm(139)$\\
\hline
\end{tabular}
\caption{For each of the layer groups in the first column,
the space groups containing the layer group as a subperiodic group are listed in the second column. Some space groups appear more than once because different two-dimensional slices can have different symmetries. 
The layer groups above the double line are symmorphic, while those below are nonsymmorphic; this determines the EBR analysis in Sec.~\ref{sec:EBR}.
However, a symmorphic layer group can be a subgroup of a nonsymmorphic group and vice versa.
Cubic groups are omitted because they do not permit a layered structure and thus the mostly-2D analysis in this work is not likely to apply.
The data is obtained from the SECTIONS application on the BCS.\cite{Aroyo2006,Aroyo2006b,Aroyo2011}}
\label{tab:spacegroups}
\end{table}

\section{Material Realizations}
\label{sec:materials}

\begin{figure*}[t]
\centering
\includegraphics[width=.9\textwidth]{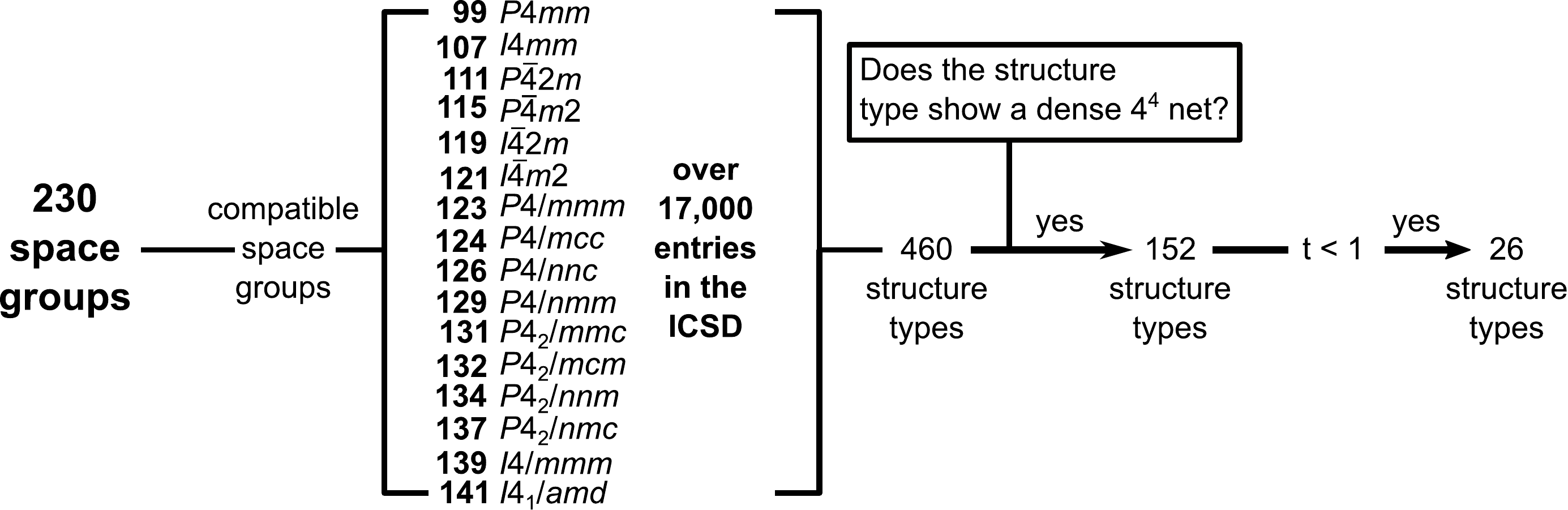}
\caption{Algorithm for finding layered square net materials that display nodal fermions. From the space groups in Table~\ref{tab:spacegroups}, we identified 460 structure types that appear with materials in the Inorganic Crystal Structure Database (ICSD)\cite{ICSD}.
Further filtering for structures containing $4^4$ nets reduced this number to 152 structure types.
Applying the tolerance factor from Ref.~\onlinecite{Klemenz2019} resulted in a final list of 26 structure types to search for promising materials.}
\label{fig:search}
\end{figure*}

Square nets, including $4^4$ nets, are common structural motifs in real materials. 
In principle, the space groups listed in Table~\ref{tab:spacegroups} can be cross-referenced with the ICSD in order to find materials with nodal points or lines in layered square net materials.
However, not all of these materials will be well described by our tight-binding model: specifically, not all materials in these space groups are layered materials (i.e., in-plane bonding is much stronger than out-of-plane bonding) and, in addition, not all materials display the $4^4$ square lattice motif.
Furthermore, since currently over
$17,000$ entries appear in the ICSD in these space groups, examining each entry individually is not feasible.
Thus, in order to find material candidates, we developed an algorithm, outlined in Fig.~\ref{fig:search}, which filters the compounds with layered $4^4$ square nets and which satisfy the tolerance factor developed by two of us in Ref.~\onlinecite{Klemenz2019}.

The tolerance factor, $t$, is defined as the ratio of interatomic distances in the $4^4$ net and the nearest neighbor atom in a different layer.
The smaller the tolerance factor, the more well separated the $4^4$ net is from the next atomic layer and the better the tight-binding model derived by folding the $4^4$ square net applies.
Klemenz et al. (Ref.~\onlinecite{Klemenz2019}) investigated the structural properties of compounds in the PbFCl structure-type family and found that the value of $t=1$ separates the topologically interesting phases ($t < 1$) from the trivial phases ($t > 1$).\cite{Klemenz2019}
Note that the tolerance factor only eliminates compounds that do not exhibit an electronic structure approximated by our tight binding model, but it does not take the Fermi level into account.
The exact position of the Fermi level is determined by the number of electrons in the $4^4$ net. The electron count in materials with $t<1$ can be between 5 and 7 electrons per net atom. 
For well-isolated $4^4$ nets, such as in ZrSiS, the band crossing points are located at the Fermi level for 6 electron systems, which correspond to half-filled $p_x$ and $p_y$ orbitals and filled $s$ and $p_z$ bands.

We now describe the algorithm depicted in Fig.~\ref{fig:search}.
The majority of the compounds in the space groups listed in Table~\ref{tab:spacegroups} have multiple entries in the ICSD, which often are repeated entries of the same compound studied at different temperatures or pressures.
In these cases, we chose the entry that represented the most precise crystal structure solution, which was obtained at standard conditions (room temperature and ambient pressure), if available.
We then checked which of the structure types exhibited a $4^4$ square-net motif.
Within the 152 structure types that occurred in the space groups in Table~\ref{tab:spacegroups} and exhibited a $4^4$ net, the unique compounds were examined with respect to the tolerance factor, \textit{t}.\cite{Klemenz2019}
Candidate Dirac materials that satisfy the tolerance factor were found in 26 of the 460 structure-types that exist in the space groups that fulfill the symmetry requirements.

In the following we describe two promising materials, ThGeSe and KCu$_2$EuTe$_4$, that came out of this search.
ThGeSe has not been previously discussed in connection with nodal fermions, while KCu$_2$EuTe$_4$ was discussed in earlier work\cite{Yang2018} but here we focus on a different aspect.
Both materials crystallize in symmorphic space groups: this reinforces the idea that nonsymmorphic symmetries are a particular route -- but not the only route -- to finding nodal fermions.
Furthermore, the nodal points are protected solely by the $p4mm$ symmetry of the $4^4$ net layer: thus, they are distinct from nodal lines that can be protected by a $z$-normal glide.
Finally, we discuss the connection to the well-known line node materials with Bi or Sb square nets.

Materials where the atoms in the dense square net reside on a $C_4$ rotation center, i.e., the left blocks of Fig.~\ref{fig:ZrSiSvariations}, did not appear in our analysis.
We conclude that this configuration is not very common in nature, likely because it is chemically unstable to have the atoms in the planes above/below the dense square net directly on top of the square net atoms.

\subsection{ThGeSe}

\begin{figure*}[t]
\centering
\includegraphics[width=.6\textwidth]{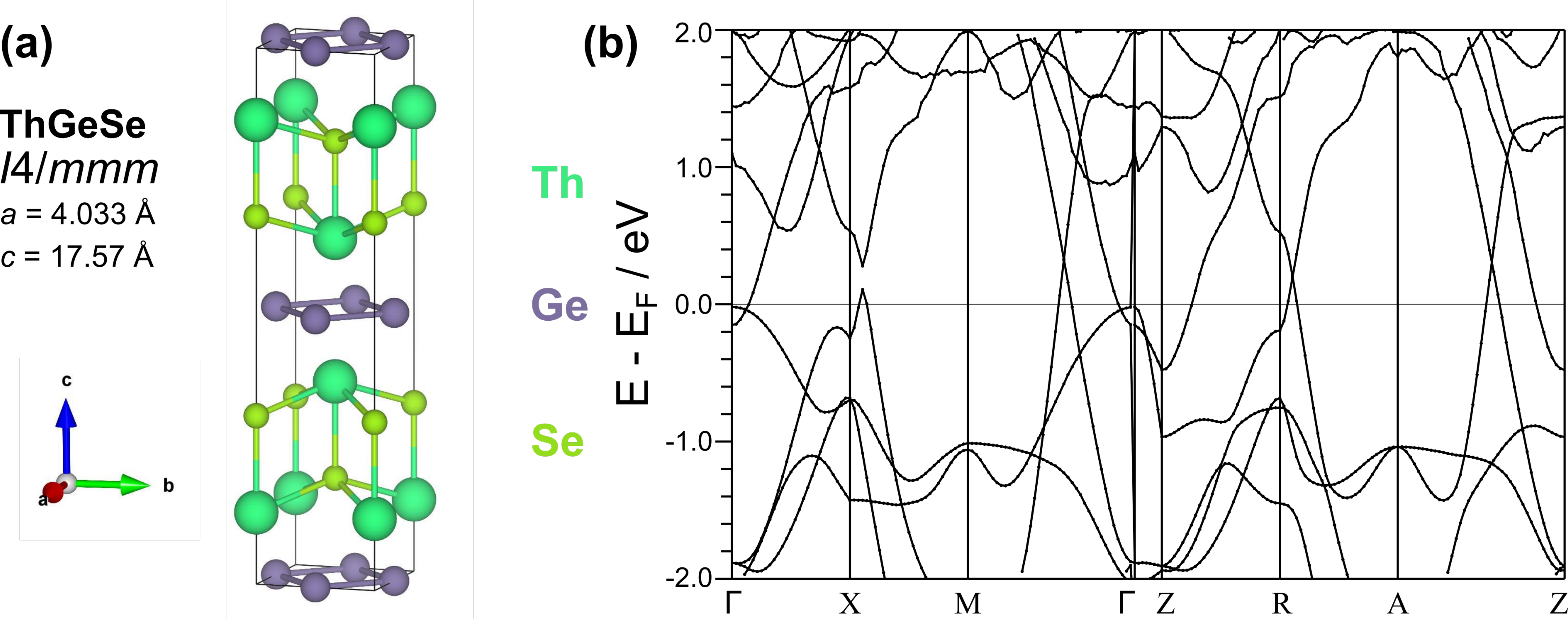}
\caption{(a) Crystal structure and (b) band structure of ThGeSe. The band structure for the body-centered tetragonal crystal is plotted with respect to the primitive tetragonal Bravais lattice for easy comparison to Fig.~\ref{fig:ZrSiSvariations}. Different colored bands represent different irreps.}
\label{fig:ThGeSe}
\end{figure*}

ThGeSe crystallizes in the space group $I4/mmm$ (139).
The ICSD structure type is named for the isostructural compound UAsTe.\cite{UAsTe} The crystal structure is very similar to that of ZrSiS, which adopts the PbFCl structure type.\cite{ZrSiS} While in ZrSiS the Si$^{2-}$ $4^4$ nets are separated by identical NaCl-like ZrS$^{2+}$ slabs, in ThGeSe the two NaCl-like slabs in the unit cell are shifted by $(1/2,1/2,0)$ relative to each other (see Fig.~\ref{fig:ThGeSe}(a)).
This difference in stacking causes the crystal to have a mirror reflection symmetry across the plane of the Ge atoms, instead of the glide symmetry present in ZrSiS.

The steep bands near the Fermi level come mostly from the Ge $p_x$ and $p_y$ orbitals (see Fig.~\ref{fig:ThGeSe}(b)), whereas the more shallow band close to the Fermi level come from Ge $p_z$ orbitals.
Further, we have computed the tolerance factor for ThGeSe, $t=0.92$. 
Hence, the Ge square net is well-separated from the neighboring Th square net and our band-folded model for layered square nets provide a starting point to understand the band structure of this materials.

Since the crystal structure shows that the Ge atoms are not located on a $C_4$ axis, we expect the nodal points to be described by the upper right block in Fig.~\ref{fig:ZrSiSvariations}: specifically, there should be Dirac points along $\Gamma-M$ and $\Gamma-X$ in the band structure in Fig.~\ref{fig:ThGeSe}(b), but not nodal lines. (Note that the band structure is plotted with respect to the primitive tetragonal Bravais lattice BZ instead of the body-centered BZ in order to make the comparison to the square lattice more clear.) 
The steep linearly dispersing bands along $\Gamma-M$ are clearly visible.
It is symmetry protected, as indicated by the fact that the different bands have different colors and therefore different symmetry eigenvalues.
This is exactly as predicted from the tight binding model.
A similar crossing along $\Gamma-X$ is not present (the crossing between green and black bands is an accidental crossing between the $p_z$ and the $p_x/p_y$ bands.)
This may be due to the fact that there is some mixing with the $d$ and $f$ orbitals in thorium that cause the energy bands of the crystal to deviate from the simplistic tight binding model.
The bands in the $k_z=\pi$ plane are very similar to those in the $k_z=0$ plane, including the Dirac crossing with steep linearly dispersing bands along $Z-A$.
This indicates the planar nature of the material.


In ThGeSe the nodal fermions are located close to the Fermi level. We understand this by assuming Th has an oxidation state of $+4$, which is reasonable for intermetallic Th compounds.
We then derive a electron distribution of Th$^{4+}$Ge$^{2-}$Se$^{2-}$, where the Ge atom has six electrons, resulting in half-filled $p_x$ and $p_y$ bands. Several thorium and uranium compounds are members of this structure type (\textit{t} values): ThGeS (0.90), ThGeTe and UGeSe (0.92), ThSiS, ThSiSe and ThSiTe (0.96), and UGeTe and USiSe (0.97), which each have six electrons in the $4^4$ net, and UAsTe (1.00) and UPTe (1.02), which each have seven electrons in the $4^4$ net. The electron counts assume that thorium will exist as Th$^{4+}$ with a 5\textit{f}$^0$ configuration and uranium as U$^{4+}$ with a 5\textit{f}$^2$ configuration.\cite{UAsTe} We expect nodal fermions to be present for all compounds with $t < 1$; the band structure will be cleanest for the smallest $t$. The Fermi level will be at the nodal point for an electron count of six electrons per net atom.

\subsection{KCu$_2$EuTe$_4$}

\begin{figure*}[t]
\centering
\includegraphics[width=.6\textwidth]{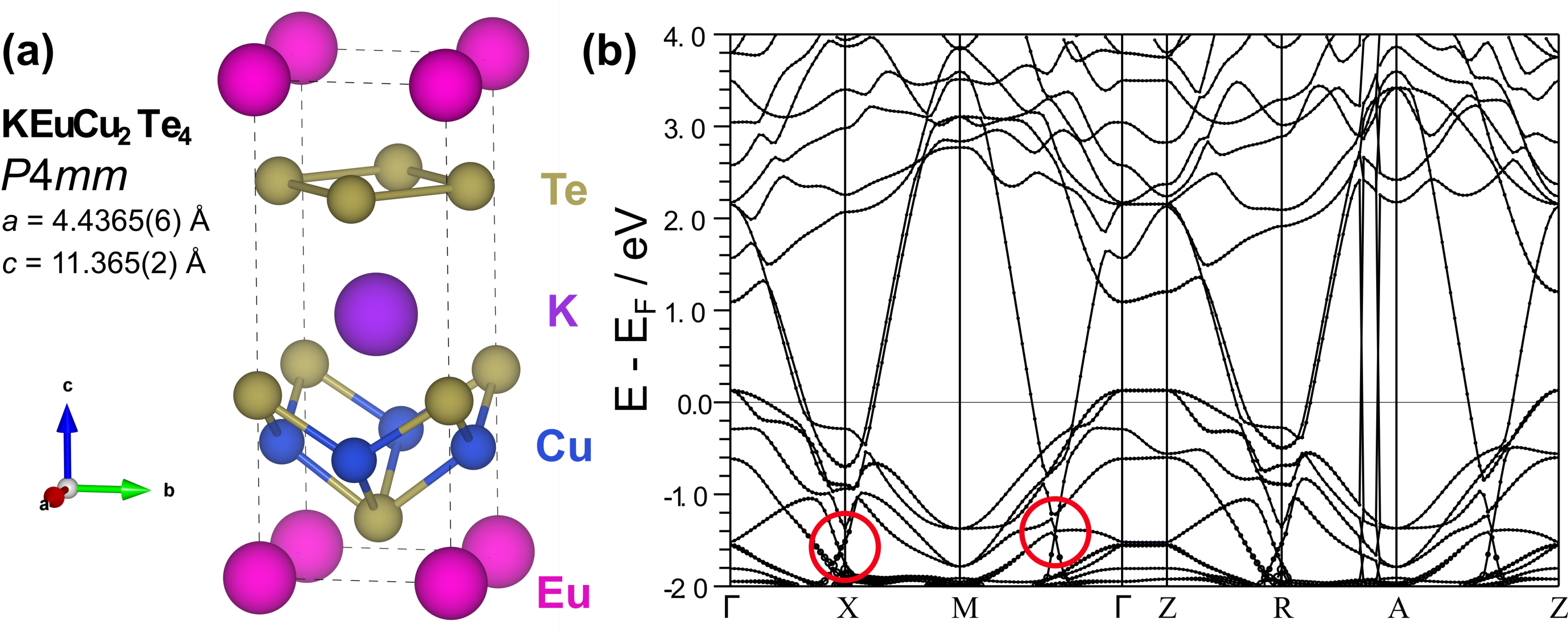}
\caption{(a) Crystal structure and (b) band structure of KCu$_2$EuTe$_4$.
The red circles around -1.5eV indicate the nodal crossings embedded in the valence bands.}
\label{fig:KEuCuTe4}
\end{figure*}

KCu$_2$EuTe$_4$\cite{KCu2EuTe4} crystallizes in the space group $P4mm$ (SG 99) and is labelled by the structure type of the same name. The material was previously reported to be a nodal line semimetal.\cite{Yang2018}
The bands near the Fermi level come from the Te $p_x$ and $p_y$ orbitals.
The crystal structure is shown in Fig.~\ref{fig:KEuCuTe4}(a).
In this structure, the Te atoms form two types of square nets. We can apply the band-folded square net model to the denser square net, which is well separated from neighboring planes of K$^+$ and Eu$^{2+}$ according to the tolerance factor $t=0.90$.
The ICSD only reports one other compound\cite{KCu2EuTe4} that exists in this structure type (Cu$_2$EuKTe$_4$), which is Na$_{0.2}$Ag$_{2.8}$EuTe$_4$, which has $t=0.93$.

Since there is no $z$-normal glide symmetry and the Te atoms in the dense square net are not a $C_4$ rotation center, the structure is described by the upper right block of Fig.~\ref{fig:ZrSiSvariations} and we expect Dirac points along the $\Gamma-M$ and $\Gamma-X$ lines, but no nodal lines.
The linearly dispersing upper half of the Dirac cones are clearly visible in the band structure in Fig.~\ref{fig:KEuCuTe4}(b): specifically, they remain linear over a range of about 1.2 eV along $\Gamma-X$ and nearly double that along $\Gamma-M$.

However, tracing the linear bands down in energy shows that the nodal point, circled in red in Fig.~\ref{fig:KEuCuTe4}(b), is located approximately 1.5eV beneath the Fermi level, due to the electron count. For KCu$_2$EuTe$_4$ the distribution of electrons can be written as K$^+$Cu$_2^+$Eu$^{2+}$Te$_2^{2-}$(Te$_2$)$^-$. 
The Te atoms in the $4^4$ net (Te$_2$)$^-$ thus have 6.5 electrons, resulting in a more than half-filled $p_x$ and $p_y$ band. 
Consequently, the Fermi level resides above the nodal points.

This conclusion relies on determining the valence state of europium, which can be ambiguous.
Lanthanides usually prefer a charge of +3.
However, previous magnetic measurements on KCu$_2$EuTe$_4$ clearly identified europium to be in the 4\textit{f}$^7$ configuration (or +2 oxidation state).\cite{KCu2EuTe4} 
If one assumes a 3+ oxidation state for Eu, the Fermi level would be located about 1 eV higher. However, this would require an electron count of more than 7 electrons per $4^4$ net atom, for which these nets become chemically unstable.\cite{a2000hypervalent} Considering that no compounds with 3+ cations in this structure type are known and that the magnetic data\cite{KCu2EuTe4} point to Eu$^{2+}$, we consider the band structure shown in Fig.~\ref{fig:KEuCuTe4}(b) to be reliable.

The band structure is nearly flat along the $\Gamma-Z$ line and the bands along $\Gamma-X-M$ are very similar to those along $Z-R-A$, and, consequently, also exhibit large linearly dispersing bands corresponding to the upper half of a Dirac cone; this further verifies treating the crystal to be a layered material.

\subsection{Bi square nets in the SmCuP$_2$ structure-type}

Our materials search also lead to many compounds that are known topological semimetals. One class is the SmCuP$_2$
structure type ($I4/mmm$). This structure type includes materials hosting anisotropic nodal fermions, such as the layered manganese pnictides,\cite{Park2011,Lee2013,Masuda2016,Liu2016,Li2016} $A$MnBi$_2$, $A=$ Sr, Ba, Eu, as well as in BaZnBi$_2$.\cite{Zhao2018} We now describe how these materials fit into the framework of the current manuscript. For this we compare these to the chemically similar compounds YbMn(Sb/Bi)$_2$\cite{Borisenko2015,Wang2016b,Kealhofer2018,Liu2017} in the HfCuSi$_2$ structure-type ($P4/nmm$).
Both structure types display Bi or Sb $4^4$ nets.

In all cases, the Bi or Sb atoms are not centers of a $C_4$ rotation.
Therefore, in the symmorphic space group ($I4/mmm$), these materials are described by the upper right block in Fig.~\ref{fig:ZrSiSvariations} and can display Dirac cones along $\Gamma-M$ and $\Gamma-X$, while in the non-symmorphic space group ($P4/nmm$), the materials are described by the lower right block in Fig.~\ref{fig:ZrSiSvariations} and can display nodal lines that cross $\Gamma-M$ and $\Gamma-X$.
All of these compounds exhibit a tolerance factor, $t$, between 0.9 and 0.93.
Since $t<1$, the inter-plane spacing exceeds the in-plane spacing and the tight-binding model describes the Bi bands well.
Consequently, nodal lines or points are apparent in the band structure, as has been previously reported.\cite{Park2011,Lee2013,Borisenko2015,Masuda2016,Liu2016,Li2016,Wang2016b,Liu2017,Kealhofer2018,Zhao2018}

\section{Elementary band representations}
\label{sec:EBR}

One of the novel aspects of the band folding procedure is that it can predict band crossings that could not be deduced from the connectivity of elementary band representations (EBRs), as long as there are no band inversions after band folding, i.e., the folded band structure qualitatively captures the relevant physics near the Fermi level.
The tolerance factor is designed to capture the crystals where this condition is likely to be satisfied ($t<1$).

The connectivity of EBRs has been computed for all space groups and is a powerful tool to predict topological semimetals and insulators.\cite{TQC,Vergniory2017,Elcoro2017,Cano2018,Bradlyn2018}
Specifically, if, for a particular material, the bands at the Fermi level transform as a ``connected'' EBR,\cite{TQC} and the orbitals are partially filled, then the material is guaranteed to be metallic.
Such a constraint cannot exist for a material where the bands near the Fermi level are derived from two EBRs.
This is because bands corresponding to two EBRs can always be realized with an energy gap: physically, the orbitals corresponding to distinct EBRs are not related by symmetry and therefore can generically have different onsite potentials and/or see different surrounding environments.

Therefore, in the cases where the orbitals corresponding to the $4^4$ square net split into two EBRs after band folding, the EBR connectivity is not enough to guarantee that the material will be a semimetal.
Instead, the assumption that band folding accurately describes the band structure (i.e., there are no bands that invert after band folding) provides the extra input necessary to guarantee that the symmetry-protected band crossings discussed in Sec.~\ref{sec:folding} are present.

Broadly speaking, in non-symmorphic groups the original EBR will not split into two EBRs, while in symmorphic groups, it will.
As an example, we compare the layer groups $p4mm$ and $p4/nmm$, which are symmorphic and non-symmorphic, respectively. 
We will show in Sec.~\ref{sec:EBRp4mm} that in $p4mm$ (symmorphic) band folding causes the original sites to split into two EBRs.
Consequently, the constraint of no band inversions after band folding is necessary to deduce the band crossings.
We then show in Sec.~\ref{sec:EBRp4nmm} that in $p4/nmm$ (non-symmorphic) the original EBR remains an EBR in the new lattice because the glide symmetry in the non-symmorphic group relates the two sites in the unit cell.
Since the EBRs with glide symmetry cannot be realized with an insulating gap (a fact that can be checked using the BANDREP application\cite{TQC,Vergniory2017,Elcoro2017} on the BCS), the band crossings cannot be removed, and the extra constraint of having no band inversions relative to band folding is unnecessary.
This is a generic feature of non-symmorphic groups.\cite{Young2015}

\subsection{Symmorphic group: EBRs in $p4mm$}
\label{sec:EBRp4mm}

\begin{figure}[h]
\centering
\begin{minipage}[t]{0.7\linewidth}
\begin{tabular}{cc}
Label & Coords\\
$1a$ & $(0,0,z)$\\
$1b$ & $(\frac{1}{2},\frac{1}{2}, z)$\\
$2c$ & $(\frac{1}{2},0,z), (0,\frac{1}{2},z)$
\end{tabular}
\end{minipage}
\begin{minipage}[t]{0.25\linewidth}
\vspace{-1.1cm}
\includegraphics[width=.9in]{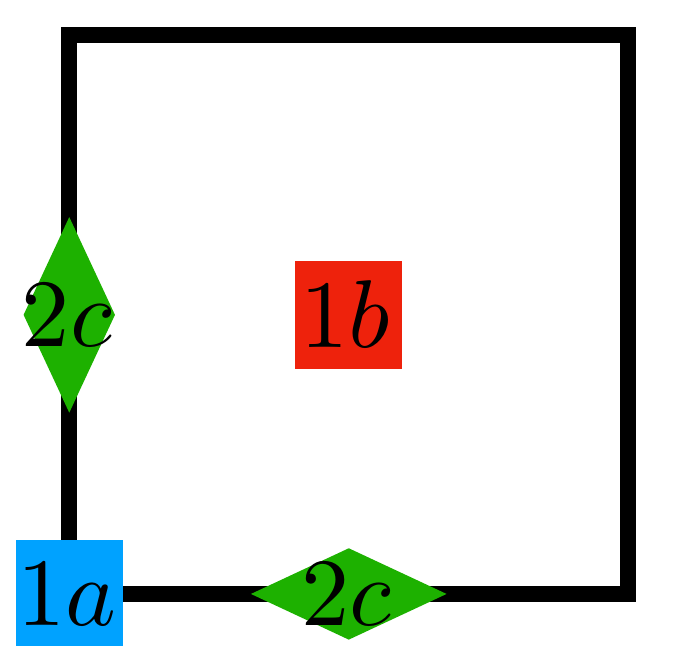}
\end{minipage}
\caption{Maximal Wyckoff positions in $p4mm$.\cite{Aroyo2006,Aroyo2006b,Aroyo2011}}
\label{fig:Wyckoffp4mm}
\end{figure}

The layer group $p4mm$ has three maximal Wyckoff positions, shown in Fig.~\ref{fig:Wyckoffp4mm}.
The $1a$ and $1b$ positions are invariant under $C_{4v}$, while sites in the $2c$ position are invariant under $C_{2v}$.

\subsubsection{Case 1: site-centered $C_4$ remains after band folding}
\label{sec:EBRp4mm1}

We first consider the center configuration in Fig.~\ref{fig:C4centers}.
Before the $\sqrt{2} \times \sqrt{2}$ unit cell is considered, the atoms reside at the $1a$ position.
When the unit cell is enlarged, the Wyckoff position splits into two positions, the $1a$ and $1b$ position.
Both are $C_4$ centers, which can be visually verified from Fig.~\ref{fig:C4centers}.
Since the site-symmetry group ($C_{4v}$) is unchanged, the orbitals on each site are still an irrep of the site-symmetry group.
Thus, the folded bands correspond to two EBRs.
Generically, two EBRs can be separated by an energy gap.
However, the constraint that no band inversions occur relative to the folded configuration guarantees that the four bands in the folded band structure are connected.
This connectivity could not be deduced from the EBRs alone.

\subsubsection{Case 2: plaquette-centered $C_4$ remains after band folding}
\label{sec:EBRp4mm2}

We now consider the left configuration in Fig.~\ref{fig:C4centers}.
Before the $\sqrt{2}\times\sqrt{2}$ unit cell is considered, the atoms reside at the $1a$ position.
After band folding, this position becomes the $2c$ position in the new unit cell.
Thus, the two sites in the enlarged unit cell are still part of the same Wyckoff position.
However, the $p_x$ and $p_y$ orbitals are no longer irreps of the site-symmetry group of the $2_c$ position (this can be easily verified since the site symmetry group, $C_{2v}$, only has one-dimensional irreps), so the orbitals each comprise a different EBR.
Thus, the folded bands again correspond to two EBRs, which can generically be separated by an energy gap.
But again, the constraint that no band inversion occur relative to the folded configuration guarantees that the four bands in the folded band structure are connected, which could not have been deduced from the EBRs alone.

\subsection{Non-symmorphic group: EBRs in $p4/nmm$}
\label{sec:EBRp4nmm}

\begin{figure}[h]
\centering
\begin{minipage}[t]{0.7\linewidth}
\begin{tabular}{cc}
Label & Coords\\
$2a$ & $(\frac{1}{2},0,0), (0,\frac{1}{2},0)$\\
$2b$ & $(0,0,z), (\frac{1}{2},\frac{1}{2}, -z)$\\
$4c$ & $(\frac{1}{4}, \frac{1}{4},0), (\frac{3}{4}, \frac{3}{4}, 0) , (\frac{1}{4},\frac{3}{4},0) , (\frac{3}{4}, \frac{1}{4},0)$
\end{tabular}
\end{minipage}
\begin{minipage}[t]{0.25\linewidth}
\vspace{-1.1cm}
\includegraphics[width=.9in]{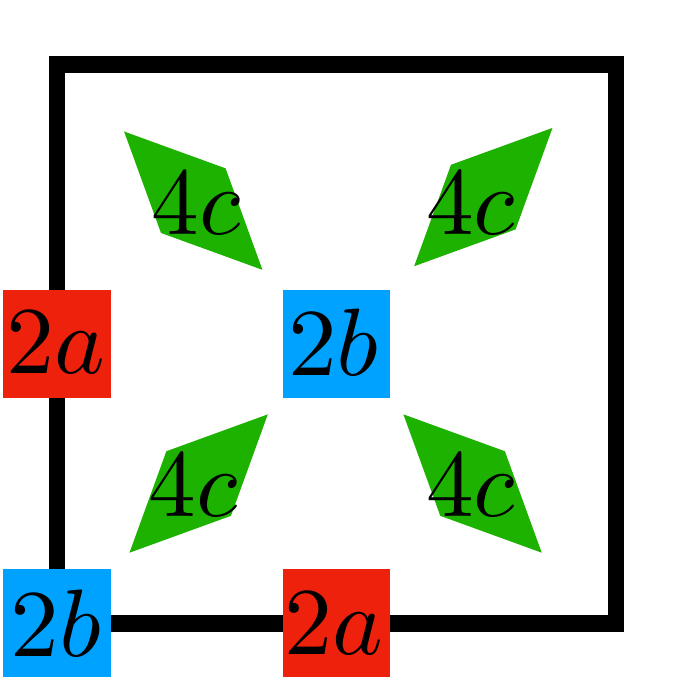}
\end{minipage}
\caption{Maximal Wyckoff positions in $p4/nmm$.\cite{Aroyo2006,Aroyo2006b,Aroyo2011} The $2a$ position is not $C_4$ invariant, but is invariant under an $S_4$ rotoinversion.}
\label{fig:Wyckoffp4nmm}
\end{figure}

For comparison, we now consider the EBRs in $p4/nmm$.
The group has three maximal Wyckoff positions, shown in Fig.~\ref{fig:Wyckoffp4nmm}.
The multiplicity of each site is always an even number because the group contains a glide symmetry.
The site-symmetry group of the $2a$ position is $D_{2d}$; the site-symmetry group of the $2b$ position is $C_{4v}$; and the site-symmetry group of the $4c$ position is $C_{2h}$.

\subsubsection{Case 1: site-centered $C_4$ remains after band folding}
\label{sec:EBRp4mnn1}

We now consider the center configuration in Fig.~\ref{fig:C4centers}. 
In the enlarged unit cell, the original atoms are $C_4$ centers, so they must be in the $2b$ position (recall the site-symmetry group of the $2a$ position, $D_{2d}$, does not have a $C_4$ center.)
Since the site-symmetry group of the $2b$ position is $C_{4v}$, the same as it was before band folding, the orbitals remain an irrep of the site-symmetry group.
Hence, they comprise a single EBR. Furthermore, utilizing the BANDREP application\cite{TQC,Vergniory2017,Elcoro2017} on the BCS shows that this EBR cannot be disconnected.
We conclude that unlike the same positions in $p4mm$ (Sec.~\ref{sec:EBRp4mm1}), the band connectivity could be deduced from the EBR connectivity (this also follows from Young and Kane.\cite{Young2015})
Consequently, the band crossings that result from band folding cannot be removed from the band structure, regardless of how the bands are deformed.

\subsubsection{Case 2: plaquette-centered $C_4$ remains after band folding}
\label{sec:EBRp4mnn2}

Finally, we consider the left configuration in Fig.~\ref{fig:C4centers}.
In the enlarged unit cell, the original atoms are not $C_4$ centers, but there are two of them in the unit cell, so we deduce that they are in the $2a$ position.
The $p_x$ and $p_y$ orbitals do transform as an irrep of the site-symmetry group ($D_{2d}$); hence, the folded bands comprise a single EBR.
Utilizing the BANDREP application\cite{TQC,Vergniory2017,Elcoro2017} on the BCS shows that this EBR cannot be disconnected.
Thus, similar to the previous case in $P4/nmm$, we conclude that the band connectivity can be deduced from the EBR connectivity and the bands must be connected.\cite{Young2015}

\section{Spin-orbit coupling}
\label{sec:soc}

We now discuss the effect of spin-orbit coupling (SOC). While SOC is nearly negligible in ZrSiS, SOC has a small effect in ThGeSe and KCu$_2$EuTe$_4$, as well as in ZrSiTe.\cite{muechler2020modular}

\begin{figure}[t]
\centering
\includegraphics[width=.47\textwidth]{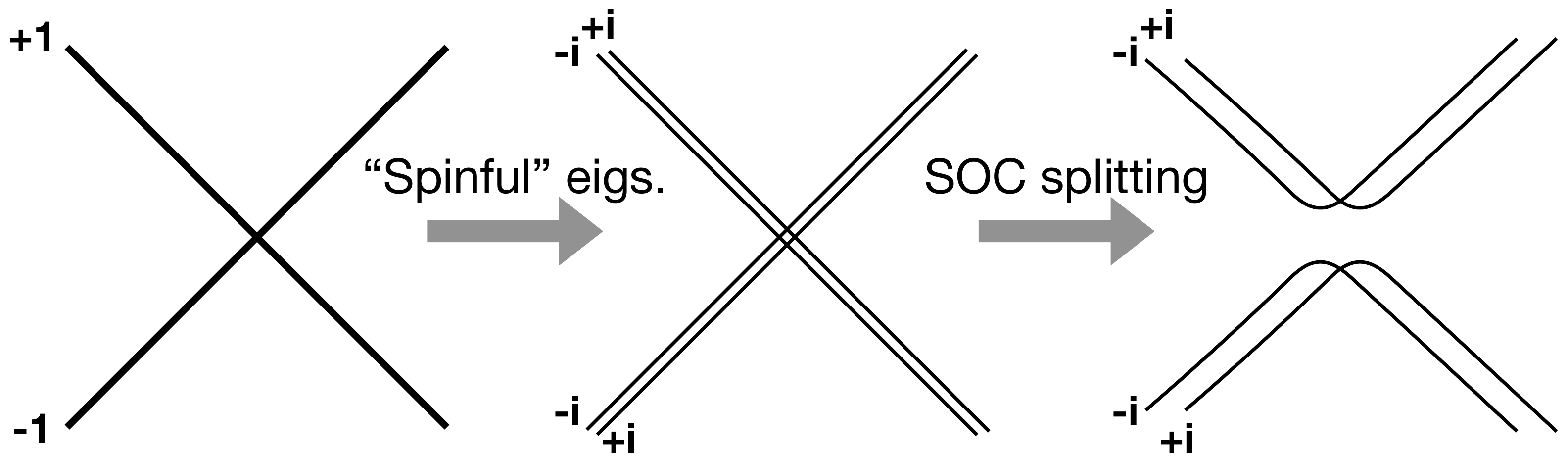}
\caption{Band crossings between spin-degenerate bands with $\pm 1$ mirror eigenvalues (left) are re-labelled with ``spinful'' $\pm i$ eigenvalues (center) to account for the action of mirror symmetry on spin. In general, bands with the same eigenvalue will gap (right).}
\label{fig:SOC}
\end{figure}

When SOC is non-negligible, it will gap the band crossings along $\Gamma-X$ and $M-\Gamma$.
One can understand this result in terms of symmetry representations:
before considering SOC, the band crossings along $\Gamma-X$ and $M-\Gamma$ are symmetry-protected because the two bands that cross have opposite mirror and/or glide eigenvalues.
When SOC is non-negligible, the symmetry representations must be modified to act on the spin degrees of freedom by a tensor product:
\begin{equation}
    \rho_o \rightarrow \rho_o \otimes \rho_{\frac{1}{2}},
\end{equation}
where $\rho_o$ and $\rho_{\frac{1}{2}}$ are the representations of the appropriate mirror operator on orbital and spin degrees of freedom, respectively.
In our case, $\rho_o = \pm 1$ is a number and $\rho_{\frac{1}{2}}$ is a $2\times 2$ matrix with $\pm i$ eigenvalues.
Thus, each spin-degenerate band without SOC will split with SOC into two bands, one with $+i$ mirror eigenvalue and the other with $-i$ mirror eigenvalue.
Since bands with the same eigenvalue will generically gap, SOC can gap the original spin-degenerate band crossing, as illustrated schematically in Fig.~\ref{fig:SOC}.

This analysis agrees with ARPES data and ab initio calculations of ZrSiTe.\cite{muechler2020modular}
Notice that the denser square net in ZrSiTe is still made of Si atoms: therefore, the effect of SOC can be non-negligible even when the denser square net is made of light atoms (Si), but the larger square net is made of heavier atoms (Te).

In non-symmorphic space groups, band crossings at the corners of the BZ may remain gapless in the presence of SOC because time-reversal symmetry can protect a four-dimensional representation.\cite{Young2015,Guan2017two,Wieder2018}
However, for the structural motif and electron filling discussed in this manuscript, these protected crossings are not at the Fermi level.

\section{Outlook}

We have studied the nodal fermions that result from embedding a dense $4^4$ square net into a larger unit cell and identified the nodal fermions that are symmetry protected in different embeddings. We provided a model that shows that nonsymmorphic symmetry is not a necessary requirement for filling constrained semimetals.
Further, some cases could not have been predicted from only the EBR connectivity.
Our analysis is specific to materials with half-filled $p_x$ and $p_y$ orbitals, but can be extended to $d$ orbitals.
In particular, since $d_{xz}$ and $d_{yz}$ orbitals transform in the same way as $p_x$ and $p_y$ orbitals (see Table~\ref{tab:C4v}), the analysis can be immediately applied to search for nodal fermions in square net materials with half-filled $d_{xz}$ and $d_{yz}$ bands. A similar analysis could be carried out for the other $d$ orbitals.

Our theoretical analysis was reinforced by a materials search that identified candidate compounds that fit our model. We introduced ThGeSe as a square net material in a symmorphic space group with Dirac nodes near the Fermi level and bands that disperse linearly over a large energy window.
We also identified several similar compounds that deserve future theoretical and experimental investigation. 
In addition, we studied the linearly dispersing bands of KEuCu$_2$Te$_4$, whose Dirac cones reside 1.5eV below the Fermi level. Finally, we introduced an algorithm (Fig.~\ref{fig:search}) that can be applied to find more Dirac materials that will be investigated in future work. 

This work thus extends previous analyses that provided filling constraints to find semimetals in nonsymmorphic space groups to symmorphic groups, for the structural motif of a $4^4$ net. 
Our results demonstrate that band-folding provides an additional route to search for material realizations of nodal fermions. 
This idea can in principle be extended beyond square nets, to other structural motifs that cause band folding.

\acknowledgements

J.C. is partially supported by the Flatiron Institute, a division of the Simons Foundation. Research at Princeton was supported by the Mabel and Arnold Beckman foundation through a Beckman Young Investigator grand awarded to L.M.S.

\bibliography{SquareNets}

\end{document}